%% file: Template.tex
\documentclass[a4paper]{article}
\usepackage{spconf,amsmath,graphicx}

\makeatletter
\newcommand{\printfnsymbol}[1]{%
  \textsuperscript{\@fnsymbol{#1}}%
}
\makeatother

\title{CNN-based Spoken Term Detection and Localization without Dynamic Programming}
\name{ Tzeviya Sylvia Fuchs\printfnsymbol{1}\thanks{\printfnsymbol{1}These authors contributed equally to this work}, Yael Segal\printfnsymbol{1}, Joseph Keshet}
\address{Bar-Ilan University, Ramat Gan, Israel\\  \texttt{\{fuchstz, segalya, jkeshet\}@cs.biu.ac.il }}

\usepackage{bbm}
\input{def}

\ninept

\begin{document}
%
\maketitle
\begin{abstract}
In this paper, we propose a spoken term detection algorithm for simultaneous prediction and localization of in-vocabulary and out-of-vocabulary terms within an audio segment. The proposed algorithm infers whether a term was uttered within a given speech signal or not by predicting the word embeddings of various parts of the speech signal and comparing them to the word embedding of the desired term. The algorithm utilizes an existing embedding space for this task and does not need to train a task-specific embedding space. At inference the algorithm simultaneously predicts all possible locations of the target term and does not need dynamic programming for optimal search. We evaluate our system on several spoken term detection tasks on read speech corpora.
\end{abstract}
\begin{keywords}
spoken term detection, event detection, speech processing, convolutional neural networks
\end{keywords}

\input{introduction}
\input{problem_setting}

\input{model}
\input{experiments}

\input{conclusions}


\bibliographystyle{IEEEbib}
\bibliography{strings,refs}

\end{document}

%% file: def.tex
\usepackage{xcolor}

\renewcommand{\P}{\ensuremath{\mathcal{P}}}

\newcommand{\x}{\ensuremath{\mathbf{x}}}
\newcommand{\e}{\ensuremath{\mathbf{e}}}
\newcommand{\f}{\ensuremath{\mathbf{f}}}

\newcommand{\seqit}[1]{\bar{#1}}
\renewcommand{\sp}{\seqit{p}}

\newcommand{\sx}{\seqit{\x}}

 \newfont{\msym}{msbm10}
\newcommand{\reals}{\mbox{\msym R}}

\DeclareMathOperator{\dissim}{dissim}

%% file: introduction.tex
\section{Introduction}

Spoken term detection (STD) is the problem of determining whether and where a target term or multi-term phrase has been uttered in a speech recording. In this setting, the target term was not necessarily given at training time. The STD problem could be defined in one of two setups. In the first setup, the term is given in a phonetic or grapheme form, and in the second setup, the term is given in an acoustic form. The second setup is referred to as ``query-by-example'' (QbE) \cite{ram2018cnn, yusuf2019low, ram2019multilingual}.



Most methods for STD are based on dynamic programming to solve the search optimization problem over an input signal or to allow duration variations of the target term using dynamic time warping \cite{8070931, prabhavalkar2012discriminative, yusuf2019empirical}. This is done by breaking the signal into time segments and then recursively finding the optimal match of each segment. More recent works use various forms of deep neural networks for the QbE problem. Some approaches define an audio-embedding space such that if the term was uttered in the input audio example, it would be projected to an embedding vector near the embedding of the audio example \cite{ram2019multilingual, multi_view}. Specifically, in \cite{multi_view}, they create a shared embedding space for both audio data and the grapheme representation of terms. Other works use Siamese networks to create the embedding space \cite{kamper2016deep, settle2016discriminative, zhu2018siamese}.


The focus of this work is to propose an alternative to the dynamic programming method by allowing a network to predict all possible occurrences of a given term. 
The proposed network operates as a multi-dimensional regressor and predicts all occurrences of the target term and its locations simultaneously. In that sense we trade the time-complexity of the dynamic programming, which is often polynomial in the segment duration and linear in the input signal duration, with the size and the depth of the network. 



This work extends the SpeechYOLO model \cite{segal2019speechyolo}. The SpeechYOLO model identifies acoustic objects in an audio signal similar to the way that objects are detected in images \cite{yolo, ssd}. The model was designed for the keyword spotting problem, where the goal was to spot and locate a predefined set of words that were given at training time. The novelty of this work is the extension of the keyword spotter to handle terms which are not seen during the training phase. 


The prediction of previously unseen terms is a challenging concept. For images, it was proposed to identify unseen objects by their similarity to the weighted sum of the confidence values of previously learned objects \cite{Demirel2018ZeroShotOD}. Such a superposition of visual objects is not applicable to spoken objects, since terms in speech cannot generally be considered as a weighted sum of other terms. Furthermore, the detection of an unseen term is challenging as it is similar to segmenting a speech signal to separate terms in an unsupervised manner. We overcome the challenge by defining unseen terms by their phonetic content and introducing them to the model using the phonetic embedding of the term.



Our spoken term detection system, which we call EmbeddingSpeechYOLO (ESY), is designed as follows: given a predefined embedding space, the model receives as input an audio file and a term in its phonetical form, which is converted to its embedding vector from the given embedding space. The ESY model then learns to predict the correct term-segmentation of the audio file it received as input. It simultaneously learns to predict the correct term embedding for every term segment it finds. Finally, its decision of whether or not the term appeared in the audio file depends on the cosine similarity between the given term-embedding and the predicted one.




This paper is organized as follows. In Section \ref{seq:problem_setting} we formally introduce the classification and localization problem setting. We  present our proposed method in Section \ref{seq:model}, and in Section \ref{sec:experiments} we show experimental results and  various applications of our proposed method. Finally, concluding remarks and future directions are discussed in Section \ref{seq:conclusions}.

%% file: problem_setting.tex
\section{Problem Setting}
\label{seq:problem_setting}

In the STD task, we are provided with a \emph{speech utterance} and a \emph{term}. The goal is to decide whether or not the term appeared within the speech utterance, and estimate its location if it was. The term is provided as a sequence of phonemes. Although the training set contains many different terms, the STD model should be able to detect any term, not only those already seen in the training phase. 


Formally, the speech signal is represented by a sequence of acoustic feature vectors $\sx  = (\x_1,\ldots,\x_T)$, where each feature vector is $D$ dimensional $\x_t\in\reals^D$ for all $1 \leq t \leq T$. A sequence of $L^q$ phonemes of a term $q$ is denoted as $\sp^q = (p_1,\ldots,p_{L^q})$, where $p_l \in \P$ for all $1 \leq l \leq L^q$ and $\P$ is the set of  phoneme symbols. We denote by $\P^*$ the set of all finite length sequences over $\P$. 

We further define a pre-trained embedding function $F:\P^* \to \reals^K$, whose input is a sequence $\sp^q$ that composes a certain term $q$. The output of the embedding function is a $K$-dimensional feature vector $\f$. The term $q$ is thus represented by the embedding vector $\f^q$.

We assume that $N$ instances of term $q$ were pronounced in the utterance $\sx$, where $N\ge 0$. Each of these $N$ events is defined by its phonetical embedding and its time location. Each such event $e$ is defined formally by the the tuple $e=(\f^q,t^q_\mathrm{start},t^q_\mathrm{end})$, where $q\in\P^*$ is the actual term that was pronounced, and $t^q_\mathrm{start}$ and $t^q_\mathrm{end}$ are its start and end times, respectively. Our goal is therefore to find all the events in an utterance, so that for each event the correct object $q$ is identified as well as its beginning and end times.




%% file: model.tex
\section{Model}
\label{seq:model}

We now describe our model formally. We assume that the input utterance $\x$ is of a fixed size $T$. We divide the input-time to $C$ non-overlapping equal sections called \emph{cells}. Each cell is in charge of detecting a single event (at most) in its time-span. That is, the $i$-th cell, denoted $c_i$, is in charge of the portion $[t_{c_i}, t_{c_{i+1}}-1]$, where $t_{c_i}$ is the start-time of the cell and $t_{c_{i+1}}-1$ is its end-time, for $1\le i \le C$. 

The cell is also in charge of localizing the detected event. The localization is defined relative to the cell's boundaries. Specifically, the location of the event is defined by the time $t \in [t_{c_i}, t_{c_{i+1}}-1]$, which is the center of the event relative to the cell's boundaries, and $\Delta t$, the  duration of the event. Note that $\Delta t$ can be longer than the time-span of the cell. Using this notation the event span is $[t_{c_i} + t-\Delta t/2 , t_{c_i} + t+\Delta t/2 ]$. 

In addition to localizing, each cell predicts an embedding vector $\f'$ that corresponds to the acoustic feature vectors appearing within the boundaries of the predicted values $t$ and $\Delta t$. The goal is for the predicted embedding vector to be an estimate of the true embedding vector of the term $q$.

The inference is performed as follows. Given a speech signal $\sx$ and a term represented as an embedding vector $\f^q$, the model predicts \emph{for each} of the $C$ cells: (i) a proposed embedding vector $\f'$; (ii) the center of the term $t'$ relative to the start-time of the cell; and (iii) its span $\Delta t'$. In order to determine whether a term $q$ has been uttered within the input audio segment, we use the cosine similarity function. If $\cos(\f^q, \f')$ is greater than a certain threshold $\phi$, we say that the term has occurred. The value of $\phi$ is tuned on the validation set.

We conclude this section by describing the training procedure. Our model, ESY, is implemented as a convolutional neural network. The initial convolution layers of the network extract features from the utterance. The output features of these layers, of size $\reals^M$, are concatenated to the input embedding vector of the term $q$ and they are both served as an input to a fully connected layer. The output of this layer is a tensor of size $C\times(K+2)$. 

Each example in the training set is the tuple $(\sx, \f^q, t^q_\mathrm{start}, t^q_\mathrm{end})$ of the acoustic signal $\sx$, the term represented by its embedding $\f^q$, and its start and end time frames, $t^q_\mathrm{start}$ and $t^q_\mathrm{end}$. Denote by $\Theta$ the set of parameters of the model. Denote by  $\mathbbm{1}^{q}_{i}$ the indicator that is 1 if the term $q$ was uttered within the cell $c_i$, and 0 otherwise. Formally, 
\begin{equation}
\mathbbm{1}^{q}_{i} = \left\{ \begin{array}{ll}
1 & t_{c_i} \le t^q  \le t_{c_{i+1}}-1 \\
0 & \mathrm{otherwise} 
\end{array} \right.~,
\end{equation}
where $t^q$ is defined as the center of event $q$. When indicating that the term is not in the cell we will use $(1-\mathbbm{1}^{q}_{i})$.

The loss function is defined as a sum over several terms, each of which takes into consideration a different aspect of the model. The first loss is in charge of improving the detection of the correct term. In our setting this is manifested by maximizing the cosine similarity between the predicted embedding and the given target embedding:
\begin{equation}
\ell_1(\sx,\f^q;\Theta) =  \sum_{i = 1}^{C} \mathbbm{1}^{q}_{i} \Big(1 - \cos(\f^q, \f')\Big) ~,
\end{equation}
where $\f'$ is the embedding predicted by the network using the parameters $\Theta$. This loss strives to achieve a cosine value between the embedding vectors which is close to 1.

The second loss function is focused on the case where the term is not uttered in a specific cell; therefore it is aimed at minimizing the similarity between the embedding vectors, for the cases where the predicted embedding does not acoustically represent the target term embedding:
\begin{equation}
\ell_2(\sx,\f^q;\Theta) =  \sum_{i = 1}^{C} (1-\mathbbm{1}^{q}_{i}) \dissim(\f^q, \f') ~,
\end{equation}




where $\dissim(\e_1,\e_2)$ is a dis-similarity function between the embedding vectors $\e_1$ and $\e_2$. We evaluated three such functions, namely, (i) $\dissim(\e_1,\e_2)=|0-\cos(\e_1,\e_2)|$,  (ii) $\dissim(\e_1,\e_2)=|-1-\cos(\e_1,\e_2)|$, or (iii) $\dissim(\e_1,\e_2)=|0-\cos^2(\e_1,\e_2)|$. The third option is more gentle and slightly reduces the dissimilarity value \cite{synnaeve2014phonetics}. 

The last loss function is aimed at finding the exact localization if the term is found in the specific cell. It has two terms: one is the mean square error between the predicted and correct term center-time relative to the corresponding cell, and the second is the mean square error between the square-root of term durations:
\begin{multline}
\ell_3(\sx,t^q_\mathrm{start}, t^q_\mathrm{end};\Theta) = \\
\sum_{i = 1}^{C}   
 \mathbbm{1}^{q}_{i} \left[ \left(t_i - t'_i\right)^2  
+ \lambda' \left( \sqrt{\Delta t_i} - \sqrt{\Delta t'_i} \right)^2 \right]~,
\end{multline}
where $t_i$ and $\Delta t_i$ are computed from $t^q_\mathrm{start}$ and $t^q_\mathrm{end}$  relative to $i$-th cell timing $t_{c_i}$. The values $t'_i$ and $\Delta t'_i$ are part of the network output with the parameters $\Theta$. 

Overall the loss function is defined as follows:
\begin{multline}
 \ell(\sx, \f^q, t^q_\mathrm{start}, t^q_\mathrm{end};\Theta) =   
 \lambda_{1} \ell_1(\sx,\f^q;\Theta) \\
 + \lambda_2 \ell_2(\sx,\f^q;\Theta)
 + \lambda_3 \ell_3(\sx,t^q_\mathrm{start}, t^q_\mathrm{end};\Theta) ~.
\end{multline}
The $\lambda$ values are used to assign different weights to the different parts of the loss function.





%% file: experiments.tex
\section{Experiments}
\label{sec:experiments}

\begin{figure}[t]
 \centering
 \includegraphics[width=\linewidth]{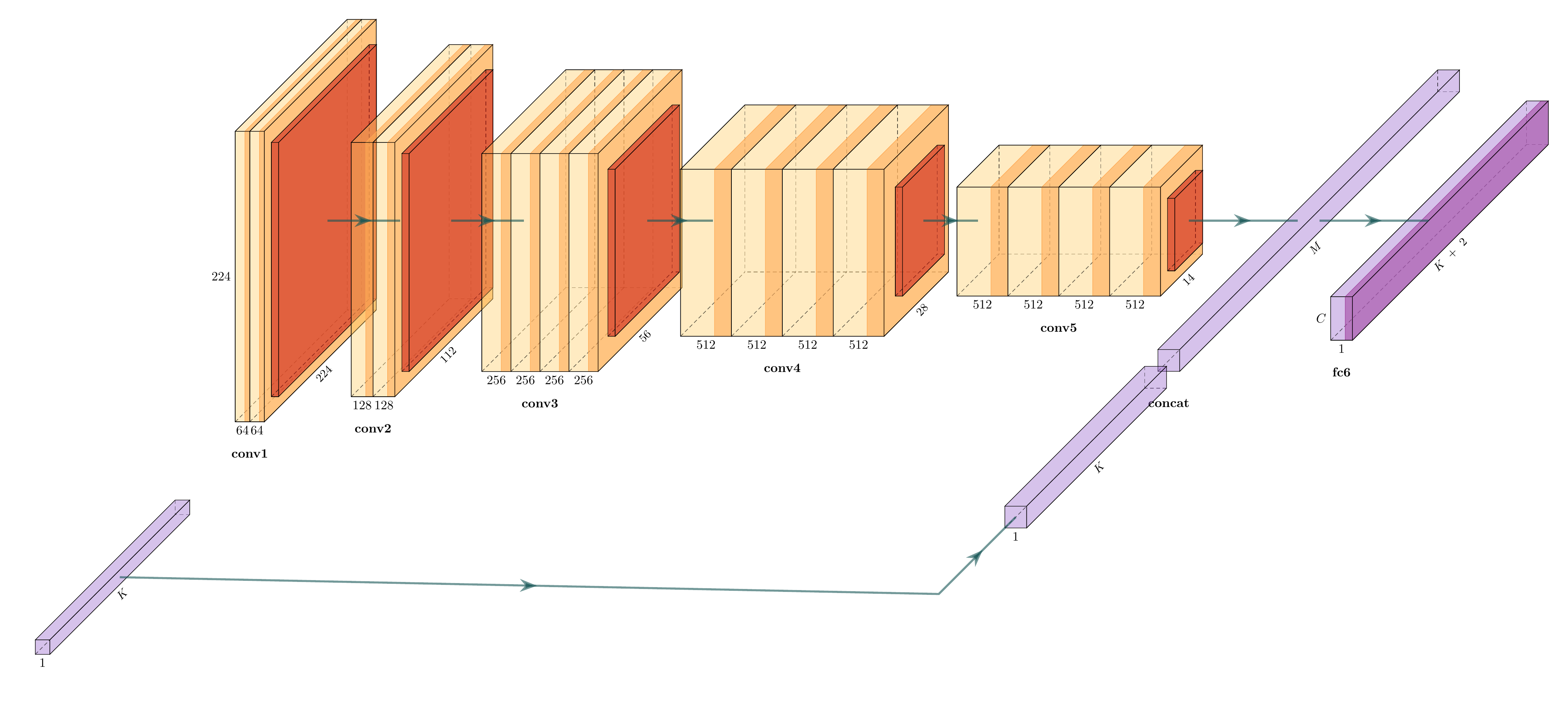}
 \caption{The detection network has 16 convolutional layers followed by concatenating the output with a given term embedding and a linear layer. Every convolutional layer is followed by BatchNorm and ReLU. }
 \label{fig:vgg19}
\end{figure}

We used a convolutional neural network that is similar to the VGG19 architecture \cite{simonyan2014very}. It had 16 convolutional layers and a fully connected layer, and the final layer predicted both event embeddings and coordinates. The model is described in detail in Fig. \ref{fig:vgg19}. It was trained using Adam \cite{kingma2014adam} and a learning rate of $10^{-3}$. We pretrained our convolutional network using the Google Command dataset \cite{warden2018speech} for  predicting $30$ predefined terms. We later replaced the last linear layer in order to perform prediction of our desired output. The size of this new final layer is $C \times (K + 2)$.

We divided our experiments into two parts: first, we evaluated ESY's prediction abilities in a dataset where a single word was articulated in every speech utterance. We then extended our experiments to speech utterances with more than one word. We evaluated our experiments using the Average Precision (AP) and Maximum Term Weight Value (MTWV) measures. 

\subsection{Datasets}

In our experiments, we use the following three datasets:

The first dataset is the Switchboard corpus of English conversational telephone speech.  We used the same setup as in \cite{multi_view}; the training set consisted of word segments whose duration ranged from 50 to 200 frames (0.5 -- 2 seconds). The train/dev/test splits contained 9971/10966/11024 pairs of acoustic segments and character sequences, corresponding to 1687/3918/3390 unique words. Mel-frequency cepstral coefficients (MFCC) features with first and second order derivatives were extracted and cepstral mean and variance normalization was applied.

The second dataset we used is  the LibriSpeech corpus \cite{panayotov2015librispeech}, which was derived from read audio books. The training set consisted of 960 hours of speech. For the test set, we used \textit{test\_clean}, which has approximately 5 hours of speech. The audio files were aligned to their given transcriptions using the Montreal Forced Aligner (MFA) \cite{mcauliffe2017montreal}. We used this dataset with two types of features: MFCC and Short-Time Fourier Transform (STFT) features. For the MFCC features, we use first and second order derivatives for the sound files, using the \texttt{librosa} package \cite{librosa}. These features were computed on a 25 ms window, with a shift of 10 ms. The STFT features were computed on a 20 ms window, with a shift of 10 ms. We define a list of search terms as follows: we randomly chose approximately 10,000 terms for training (excluding stop words). We assigned 1,500 terms for the test set, half of which are In-Vocabulary (IV) terms, i.e. they also appear in the training set, and half of which are Out-Of-Vocabulary (OOV) terms, i.e. they do not appear in the training set. We assign 1,500 terms for the validation set as well, half of which are IV and half of which are OOV.

The third dataset we used is the TIMIT corpus \cite{garofolo1993darpa}. We used this dataset for testing purposes and not for training. For this corpus, we used the MFCC features, with the aforementioned parameters. The test set was composed of 80 terms that were suggested as a benchmark in \cite{KeshetGrBe09}. For each term, we used all utterances in which the term appeared, and chose another 80 utterances in which the term has not appeared. The utterances were taken from the TIMIT test set.

\subsection{STD with a single word setup}

In this part a single word appears in every speech utterance. For these experiments, we use the Switchboard dataset. Since every audio file contains a single word, we set the number of cells $C = 1$. The embedding size is set to be $K = 1024$. The loss's $\lambda$ values are $\lambda_1 = 0.5$, $\lambda_2 = 1$ and $\lambda_3 = 2$, and were tuned on the validation set.

We obtain the embedding function $F$ by implementing \cite{multi_view} for acoustic word embeddings. We refer to their work as \textit{Multi-View}. In their work, they train two LSTM networks; one receives an audio signal, and the other receives a sequence of letters corresponding to a term that has or has not been uttered within the audio signal. Their goal is to bring the LSTMs' outputs to produce embedding vectors. They use a contrastive loss function, which is designed to make the distance between the LSTMs' outputs smaller if the input term has been uttered in the audio signal, than if the term has not been uttered in the audio signal. We denote the embedding vectors created by the \textit{Multi-View} algorithm as \emph{MV} vectors. Thus, we assign \textit{Multi-View} to create our embedding space from which the embedding vectors $\f^q$ will be drawn - with one difference:  we used phonemes to represent our term, instead of letters. This is due to the fact that phonemes are better than graphemes for representing the sound of a term. If there were several ways to pronounce a term, we used the average of the embedding vectors of all pronunciations.

We compare our work to an STD system implemented by \cite{multi_view}, using the  aforementioned \textit{Multi-View} algorithm. We also compare to  \cite{kamper2016deep} and \cite{settle2016discriminative}, in which the authors present an STD system for a single word setup, using a Siamese CNN and a Siamese LSTM, respectively. Results are shown in Table \ref{switchboard}.

 In \cite{kamper2016deep} and \cite{settle2016discriminative}, the reference term is an acoustic signal (an acoustic-to-acoustic setup), as opposed to \cite{multi_view} and our work (an acoustic-to-text setup). In this context, we find the acoustic-to-acoustic setup to be very similar to the acoustic-to-text setup, since in both cases the predictions are based on the embedding vectors' similarity, and are independent of the reference term's form.

As seen in Table \ref{switchboard}, the \textit{Multi-View} algorithm achieved better results than those of the \textit{ESY} algorithm. Note that  \textit{ESY} uses the \textit{MV} embedding vectors, each representing a single word, as its target.  Therefore, the STD capabilities of \textit{ESY} are bounded by those of \textit{Multi-View}, but are still sufficiently high. Furthermore, its AP values are significantly higher than those of \cite{kamper2016deep} and \cite{settle2016discriminative}.



\begin{table}
\renewcommand{\arraystretch}{1}
  \centering
  \resizebox{\columnwidth}{!}{
  \begin{tabular}{l c c}
    \hline
   \multicolumn{1}{c}{} & \multicolumn{1}{c}{\textbf{Acoustic-to-Text AP}} & 
   \multicolumn{1}{c}{\textbf{Acoustic-to-Acoustic AP}} \\
    \hline
    \textit{Siamese CNN} \cite{kamper2016deep} & & 0.549* \\
    \textit{Siamese LSTM} \cite{settle2016discriminative} & & 0.671* \\
    \textit{Multi-View} & \textbf{0.897} & 0.806* \\
    \textit{ESY} &  0.835 &  \\
    \hline
  \end{tabular}}
 \caption{ \label{switchboard} AP measure for several single-word STD algorithms on Switchboard test set.  In the Acoustic-to-Text setup, the term is given in textual form, whereas in the Acoustic-to-Acoustic setup, the term is given as an acoustic signal. *results reported from paper. }
\end{table}

\subsection{STD with a multiple words setup}


In this part several words appear in every speech utterance. For these experiments, we use the Librispeech dataset. We set the number of cells $C = 3$, to cover the possibility of several words occurring in the speech signal. The embedding size is set to be $K = 1024$, as before. The loss's $\lambda$ values are $\lambda_1 = 1$, $\lambda_2 = 0.5$ and $\lambda_3 = 3$, and were tuned on the validation set.

To obtain the embedding vectors of our chosen terms, we trained the \textit{Multi-View} algorithm on the LibriSpeech-360 dataset, which we segmented into single words, thus creating \emph{MV} vectors.  We then trained \textit{ESY} on audio files that were 1.5 seconds long and containing several words, while using these \emph{MV} vectors as the representation $\f^q$ of the term $q$ that was being searched for.

Table \ref{couscous} presents the localization and prediction results of our algorithm with the various versions of the dis-similarity function mentioned in Section \ref{seq:model}. We measure localization using the Intersection-Over-Union (IOU) metric, which has been calculated only on instances that have been correctly detected. Our algorithm's IOU values are exceptionally high in all settings. We find that it performed surprisingly well even on the OOV terms.

We measure our prediction ability using the AP metric. The results in Table \ref{couscous} show that the $|0-\cos^2(\e_1,\e_2)|$  dis-similarity function gave the best results. We believe that this is due to its soft correction of mistakes. Note that the dis-similarity function is used in the loss function when the term did not occur in the given cell. However, we have no knowledge of what \emph{has} been uttered in that cell; it may have been a term acoustically similar to the search term. Therefore, in such a case, a softer penalty is more appropriate.

%

\begin{table}
\renewcommand{\arraystretch}{1}
  \centering
  \resizebox{\columnwidth}{!}{
  \begin{tabular}{l c c c c}
    \hline
   \multicolumn{1}{c}{$\dissim$ function} & \multicolumn{2}{c}{\textbf{In-Vocabulary (IV)}} & 
   \multicolumn{2}{c}{\textbf{Out-Of-Vocabulary(OOV)}} \\
   \multicolumn{1}{c}{} & \multicolumn{1}{c}{\textbf{AP}} & 
   \multicolumn{1}{c}{\textbf{IOU}} & \multicolumn{1}{c}{\textbf{AP}} & 
   \multicolumn{1}{c}{\textbf{IOU}} \\
    \hline
    $|0-\cos(\e_1,\e_2)|$ &  0.679 & 0.846 & 0.1 &  0.677  \\
    $|-1-\cos(\e_1,\e_2)|$ &  0.498 & 0.805 & 0.03 &  0.596 \\
    $|0-\cos^2(\e_1,\e_2)|$  & \textbf{0.72} &  \textbf{0.852}   & \textbf{0.18} &  \textbf{0.697} \\
    \hline
  \end{tabular}}
  \caption{\label{couscous}AP and IOU values for ESY using the LibriSpeech dataset, for three proposed dis-similarity functions. }
\end{table}

To compare our algorithm's performance, we adapted the \emph{Multi-View} STD system from \cite{multi_view} to match the multiple words setup. We were inspired by previous works for QbE \cite{settle2017query, levin2015segmental}; the  QbE algorithm in \cite{settle2017query} learned single-word embeddings using some embedding function $F$; for evaluation, it then used $F$ to embed the term and the full speech utterance (which consisted of several words). The two embeddings were then compared in order to decide whether or not the term occurred within the speech utterance. Accordingly, and for fair comparison, we trained the \emph{Multi-View} model to create \emph{MV} embedding vectors, using the single-word setup on the LibriSpeech dataset. Then for evaluation, we used the same model to embed both a single term and a corresponding full speech utterance (of length 1.5 seconds), and compared the two embeddings. As opposed to \cite{settle2017query}, in our case the term is embedded by its phonetical form  and not by its acoustic form. We denote this setup as \textit{Multi-View-QbE}, as it was inspired by a QbE system. 

Further, we used an additional version of the \emph{MV} vectors, where as opposed to the \textit{Multi-View-QbE} setup, this time they were \emph{trained} in a multi-word setup. Specifically, they were trained on 1.5 second long segments, where each segment could contain several words. (This is different from their paper's original setup \cite{multi_view}, where the algorithm searched for a term in an audio segment of a single word). We denote this system as \textit{Multi-View-STD}. 

Results for the AP measure for both \textit{Multi-View-QbE} and \textit{Multi-View-STD} are presented in Table \ref{librispeech}. The \textit{ESY} algorithm achieved best results. The \textit{Multi-View-STD} algorithm, which was trained specifically on multi-word segments performed better than Multi-View-QbE, which was trained on single-word segments. Notice that in this case, \textit{ESY} is not bounded by the performance of \textit{Multi-View} since the speech utterances contain more than one word. We further present MTWV results for \textit{Multi-View-STD} and \textit{ESY} in Table \ref{librispeech_mtwv}. The \textit{ESY} algorithm once again achieved better results, this time when using the STFT features. 

\begin{table}
  \centering
  \resizebox{\columnwidth}{!}{
  \begin{tabular}{@{\extracolsep{4pt}}lccccc}
  \hline
    &  \textbf{In-Vocabulary (IV)} & \textbf{Out-Of-Vocabulary (OOV)} \\
    \hline
    \textit{Multi-View-QbE} (MFCC) &  0.003 & 0.003 \\
    \textit{Multi-View-STD} (MFCC) & 0.245 & 0.06\\
    \textit{ESY} (MFCC) & \textbf{0.72} & \textbf{0.18} \\
     \hline
     \textit{Multi-View-QbE} (STFT) &  0.002 & 0.003 \\
    \textit{Multi-View-STD} (STFT) & 0.09  & 0.028 \\
    \textit{ESY} (STFT) & \textbf{0.71} & \textbf{0.17}  \\
    \hline
  \end{tabular}}
   \caption{\label{librispeech} AP evaluation of \textit{ESY} vs. \textit{Multi-View} on LibriSpeech dataset. \textit{ESY} results used $|0-\cos^2(\e_1,\e_2)|$ dis-similarity function.}
\end{table}



\begin{table}
  \centering
  \resizebox{\columnwidth}{!}{
  \begin{tabular}{@{\extracolsep{4pt}}lccccc}
  \hline
    &  \textbf{In-Vocabulary (IV)} & \textbf{Out-Of-Vocabulary (OOV)} \\
    \hline
    \textit{Multi-View-STD} (MFCC) & 0.010  & 0.005 \\
    \textit{ESY} (MFCC) & \textbf{0.330}  & \textbf{0.136}  \\
    \hline
    \textit{Multi-View-STD} (STFT) & 0.000  & 0.000 \\
    \textit{ESY} (STFT) & \textbf{0.348} & \textbf{0.142}  \\
    \hline
  \end{tabular}}
  \caption{\label{librispeech_mtwv} MTWV evaluation of \textit{ESY} vs. \textit{Multi-View} on LibriSpeech dataset. \textit{ESY} results used $|0-\cos^2(\e_1,\e_2)|$ dis-similarity function.}
\end{table}

In our previous experiments, the dataset that the \textit{ESY} algorithm was tested on was from the same distribution as the dataset with which we trained our embedding function $F$. Therefore, we tested our model using the TIMIT dataset's test set. The majority of the TIMIT's selected terms did not appear in the LibriSpeech training set. We use the \textit{ESY} model that has been trained on the LibriSpeech dataset with MFCC features. We received an AP value of 0.606. TIMIT's results are better than those achieved on LibriSpeech in the OOV setup; this could be explained by the fact that TIMIT's dataset is a cleaner dataset. Nevertheless, it proves that our model achieves satisfactory results even when using a dataset that is from a distribution other than the one we trained on.

In addition, we compare \textit{ESY}'s localization capabilities with those of MFA. To do so, we train the MFA and \textit{ESY} algorithm on LibriSpeech, and test both models on the test set of the TIMIT corpus. The forced aligner, MFA, uses a complete transcription of the words uttered in the audio file to perform its alignments. This is in contrast to the \textit{ESY} algorithm, which receives no information about the terms uttered within the given speech signal. Therefore, due to the fact that MFA receives additional textual information, we considered its localization ability to be an ``upper bound'' to ours. The IOU measure was used to compare both algorithms' output alignments with TIMIT's given word alignments. The IOU of \textit{ESY} on TIMIT was 0.73, while MFA achieved 0.9. Thus, we found that \textit{ESY}'s IOU value, while lower than MFA's, was sufficiently high.

%% file: conclusions.tex
\section{Conclusions}
\label{seq:conclusions}

In this work we presented an end-to-end system for detection and localization of textual terms. We evaluate our system on the spoken term detection task for both single-word and multiple-words setups. We further present results regarding localization. Future work includes improving the AP values for the OOV terms, and combining the SpeechYOLO algorithm for IV terms with the \textit{ESY} algorithm for OOV terms.

\section{Acknowledgements}

T. S. Fuchs is sponsored by the Malag scholarship for outstanding doctoral students in the high tech professions. Y. Segal is sponsored by the Ministry of Science \& Technology, Israel. The authors would like to thank Sharadhi Alape Suryanarayana for helpful comments on this paper.